# Denoise Stepwise Signals by Diffusion Model Based Approach

Xingdi Tong[1], Chenyu Wen[2*]

1. Department of Civil and Industrial Engineering, Ångströmlaboratoriet, Uppsala University, Lägerhyddsvägen 1, 75237, Uppsala, Sweden

2. Division of Solid-State Electronics, Department of Electrical Engineering, Ångströmlaboratoriet, Uppsala University, Lägerhyddsvägen 1, 75237, Uppsala, Sweden

* Corresponding author: Chenyu Wen: chenyu.wen@angstrom.uu.se

**Abstract:**

Stepwise signals are ubiquitous in single-molecule detections, where abrupt changes in signal levels typically correspond to molecular conformational changes or state transitions. However, these features are inevitably obscured by noise, leading to uncertainty in estimating both signal levels and transition points. Traditional frequency-domain filtering is ineffective for denoising stepwise signals, as edge-related high-frequency components strongly overlap with noise. Although Hidden Markov Model–based approaches are widely used, they rely on stationarity assumptions and are not specifically designed for signal denoising. Here, we propose a diffusion model-based algorithm for stepwise signal denoising, named the Stepwise Signal Diffusion Model (SSDM). During training, SSDM learns the statistical structure of stepwise signals via a forward diffusion process that progressively adds noise. In the following reverse process, the model reconstructs clean signals from noisy observations, integrating a multi-scale convolutional network with an attention mechanism. Training data are generated by simulating stepwise signals through a Markov process with additive Gaussian noise. Across a broad range of signal-to-noise ratios, SSDM consistently outperforms traditional methods in both signal level reconstruction and transition point detection. Its effectiveness is further demonstrated on experimental data from single-molecule Förster Resonance Energy Transfer and nanopore DNA translocation measurements. Overall, SSDM provides a general and robust framework for recovering stepwise signals in various single-molecule detections and other physical systems exhibiting discrete state transitions.

## 1. Introduction

Stepwise signals widely exist in various single molecule detections, in which target molecules are traced and interrogated one by one. Thus, their signals usually show stepwise shifting among different levels, instead of smoothly varying by the average effect of $10^{10}$-$10^{23}$ (pM-M) molecules in ensemble measurements. Single molecule detection technologies vary from electrical signal-based manners, including nanopore DNA and protein sequencing[1–4], translocation[5,6], and trapping[7–9] to optical signal-based manners, such as single-molecule Förster Resonance Energy Transfer (sm-FRET)[10,11], plasmonic probes[12], and single molecule fluorescence[13,14], and to mechanical signal-based manners, such as atomic force microscope[15,16] and optical/magnetic tweezers[17,18] The stepwise shifting among different levels indicates the transition of individual molecules among several micro states[19]. Thus, the statistical features of these steps, such as the transition time point, step height, and interval between transitions, reflect molecule structures, dynamic behaviors, and interaction mechanisms[20,21]. As can be anticipated, low signal-to-noise ratio (SNR) raises a major challenge in processing the stepwise signals. Severe background noise from, for example, thermal vibration[22], electromagnetic interference[23], and instrumental noise[24], leads to distortion and blurred step edges, hindering analysis accuracy[25]. Therefore, denoising is crucial for achieving accurate single-molecule recognition, state transition detection, and kinetic parameter extraction.

Traditional frequency domain filtering methods, such as low-pass filters, are effective only when the noise frequency components are clearly separated from those of the signal. For stepwise signals, however, transition edges contain abundant high-frequency components that strongly overlap with the high-frequency noise[26,27], forcing an unavoidable trade-off between noise suppression and transition edge preservation. Moreover, they often induce delay and attenuated amplitude, distorting the signal features. Several approaches based on statistical characteristics of signals have been proposed to address these limitations. Independent component analysis (ICA) was adopted for denoising nanopore translocation signals by exploiting correlations between current signal and the noise from the corresponding voltage stimuli[28]. However, its performance relies on strong current–voltage noise correlation, which is not always guaranteed. Cumulative sums algorithm (CUSUM) detects steps by modelling the signal as a piecewise constant function and monitoring the cumulative deviation of the signal[29]. However, it struggles with short events and requires user-defined parameters.

Hidden Markov Model (HMM)-based approaches are widely used to extract stochastic process parameters, such as state transition probabilities and dwell times[30,31]. However, they assume a stationary Markov process [32,33], which is not always true in reality. In addition, the results are sensitive to noise[34], since they are not specially designed for denoising. Practically, HMMs require a predefined number of hidden states. To guess the state number, additional trial-and-error-like mechanisms are introduced [35–37], increasing computational cost [38] and reliance on manual parameter tuning. *AutoStepfinder* provides automatic step detection by iteratively minimizing the difference between the data and the fitted segments[39]. However, its performance degrades under high noise conditions and, like HMMs, focuses on step extraction rather than denoising.

Recently, machine learning approaches have shown promise in stepwise signal processing. Nano Tree employs decision trees and the AdaBoost mechanism to fit steps in the signals[40]. Its accuracy decreases for signals with gradual transitions. Neural network-based methods, such as convolutional autoencoder integrated in a U-Net architecture, leverage powerful feature extraction to reconstruct noise-free signals[41]. Nevertheless, under low SNRs, it remains challenging to achieve both signal smoothness and precise detection of transition points[42].

Given this context, to denoise stepwise signals, a natural choice is a method that does not rely on explicit state modelling and can directly learn the stochastic relationship between noise and signal from data. Denoising Diffusion Probabilistic Models (DDPM), as an emerging generative framework, is such option[43]. The model emulates thermodynamic diffusion process[44] and has achieved promising performance in denoising images, audio, and temporal signals [45–47].

Here we propose a diffusion-model based algorithm for stepwise signal denoising, named Stepwise Signal Denoising Model, SSDM. Compared to aforementioned approaches, SSDM neither requires a clear distinction of signals and noise in frequency spectrum, nor a predefined number of states. As far as we know, this is the first time to develop a diffusion model-based algorithm for denoising of stepwise signal. We train the SSDM on the artificially generated noisy signals by a Markov process with known additive Gaussian noise. The denoising performance is systematically evaluated under different SNR conditions. To benchmark, we quantitatively compare the denoising results with traditional filtering methods and HMM. The comparison focuses on noise suppression capability and the accuracy of transition point recognition. Finally, to further evaluate the effectiveness and generalization capacity, we applied SSDM to experimental data from optical sm-FRET and electrical nanopore λ-DNA translocation measurements.

## 2. Methods
### 2.1. Stepwise Single Diffusion Model

We adopt DDPM as the framework of SSDM, which involves forward and reverse diffusion processes, as illustrated in Figure 1 (a). In the forward process, a Gaussian noise is gradually injected into the original noise-free signal $x_0$ until the signal deforms into an extremely noisy signal $x_T$. In the reverse process, the objective of SSDM is to predict the added noise and to recover the noise-free signal. This reverse process is defined as a Markov chain where the conditional probability from time step $t$ to $t$-1 is modeled as a Gaussian distribution parameterized by $\theta$.

$$p_\theta(x_{t-1}|x_t) \coloneqq \mathcal{N}\big(x_{t-1}; \mu_\theta(x_t, t), \Sigma_\theta(x_t, t)\big) \tag{1}$$

where, $x_t$ is the noisy observation at time step $t$, $x_{t-1}$ is the predicted value at time step $t$-1, $\mu_\theta(x_t, t)$ is the mean parameter predicted by the neural network, $\Sigma_\theta(x_t, t)$ is the variance of the Gaussian distribution. $\mu_\theta$ can be reparameterized as a function of the input $x_t$ and the noise $\epsilon$, so we do not employ the neural network to predict directly[43]. Our neural network is designed as a function approximator to predict the noise $\epsilon_\theta(x_t, t)$ injected at each step $t$, and the mean will be derived as:

$$\mu_\theta(x_t, t) = \frac{1}{\sqrt{\alpha_t}} \left( x_t - \frac{\beta_t}{\sqrt{1-\overline{\alpha}_t}} \epsilon_\theta(x_t, t) \right) \quad (2)$$

with $\beta_t = \min\left(1 - \frac{f(t)}{f(t-1)}, 0.999\right)$, $f(t) = \cos^2\left(\frac{t/T+s}{1+s} \cdot \frac{\pi}{2}\right)$, $s = 0.008$, and $T = 1000$. $\beta_t$ is the noise schedule which employs cosine schedule. $\alpha_t = 1-\beta_t$ and $\overline{\alpha_t} = \prod_{s=1}^{t} \alpha_s$ reflecting the cumulative proportion of the signal preserved during diffusion. Regarding the variance $\Sigma_\theta(x_t, t)$, we set it to a fixed schedule $\beta_t I$, where $I$ is the identity matrix. This simplifies the sampling process by directly using the noise schedule parameters.

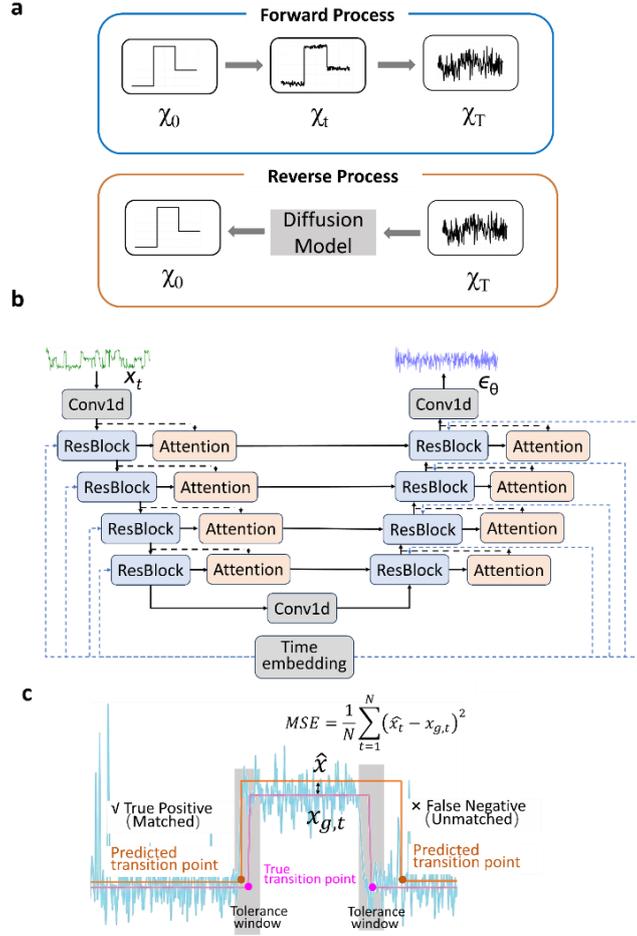

Figure 1. Overall structure of SSDM and the performance evaluation. (a) Schematics showing the forward and reverse diffusion processes of SSDM. (b) SSDM network architecture. (c) Schematics illustrating the performance evaluation metrics, *F*1 and *MSE*.

## 2.2 Network Architecture

To denoise stepwise signals, we need a deep learning model that can capture long-range dependency while preserving local sudden changes. The multi-scale architecture of U-Net is suitable for modelling this type of data, as shown in Figure 1 (b). We designed a one-dimensional (1D) convolutional neural network based on the U-Net architecture to predict noise $\epsilon_\theta(x_t, t)$. This network adopts an encoder-decoder structure and fuses multi-scale features through skip connections. Input is a 1D noisy stepwise signal $x_t$ with shape 1×1000. The embedding vector at diffusion time step $t$ is also fed into the network, enabling the model to perform noise prediction at different SNR conditions. In the encoder process, the network performs a downsampling on signal at each layer, thereby increasing the temporal scope captured by subsequent convolutions. The decoder performs upsampling layer by layer with a symmetric structure as the encoder. Along the flow, it merges the corresponding encoder features into the decoding path through skip connections at each layer. This approach prevents detail loss caused by downsampling, enabling the network to simultaneously learn local fine structures and global general patterns. Each encoder and decoder layer is implemented using residual block (Resblock) composed of convolution, normalization, activation, and skip connections. For the specific Resblock architecture, refer to Figure S1 in the Supplementary Information (SI). Convolutional structures can efficiently capture local patterns but have limited capacity to model global dependencies across time steps. To address this issue, an Attention Block (refer to Figure S2 in the SI) was connected after the ResBlock in each layer of both the encoder and decoder. The details of the entire network architecture can be found in Supporting Note 1 in the SI.

### 2.3. Data preparation

To train the model, we simulated stepwise signals by a 1D continuous time Markov chain. By controlling the transfer rate matrix, the number of states, SNR, and the amplitude level, we generated a reproducible, scalable, and representative training dataset (see Supporting Note 2 in the SI for details). The dataset comprised balanced 2-, 3-, and 4-state signals, each 1,000 time points with a normalized sampling interval. For each parameter combination, 100 independent trajectories were generated. In total, the training set contained 10,800 noise-free signals, including 2,400 2-state signals from 8 transition rate matrices, 3,600 3-state signals from 12 transition rate matrices, and 4,800 4-state signals from 16 transition rate matrices.

Noisy signals were constructed by adding Gaussian noise to the generated signals. SNR is defined as the ratio of the minimum inter-level difference of the signal to the peak-to-peak amplitude of noise (estimated as 6 times its root-mean-square value). Details of noise generated can be found in Supporting Note 3 in the SI. An independent test dataset was generated to evaluate generalization and robustness. It consisted of 3,600 signals with faster transitions, spanning SNRs of 0.25, 0.5, 1, 3, and 5. The test set included 40, 80, and 80 different parameter combinations for 2-, 3-, and 4-state signals, respectively, with 20 trajectories per combination. To investigate the performance of SSDM with known number of states, we generated separate datasets containing only 2-, 3-, or 4-state signals for training and test, while keeping dataset size and SNR range identical.

The experimental single-molecule detection data are from two platforms: sm-FRET optical signals and λ-DNA nanopore translocation electrical signals. The sm-FRET dataset[10] consists of 19 experimental trajectories with 100 Hz sampling rate, in total 226,100 data points. The λ-DNA nanopore dataset [48] comprises ionic current traces recorded at different bias voltages, with 78 pM λ-DNA dispersed in a 500 mM KCl solution, and sampled at 10 kHz. All experimental signals were normalized before denoising and rescaled to their original amplitudes and temporal scales for analysis.

### 2.4. Model training

The core objective of training is to enable the network to predict the noise at each step, $t$. Training follows a supervised learning procedure, with labels generated in real time during the forward diffusion process. We set the total number of diffusion steps to $T = 1000$ to ensure sufficient temporal resolution for smoothly transforming noise-free signals into pure noise. To improve training efficiency, we employed importance sampling over diffusion steps. Specially, over time steps $t \in [0, T]$ were sampled from a non-uniform distribution $p(t) \propto \exp(-3t/T)$, which preferentially samples smaller $t$. Based on the Cosine Schedule, $\overline{\alpha_t}$ was calculated and the noisy sample $x_t$ for the current time step was generated by adding noise $\epsilon$:

$$x_t = \sqrt{\overline{\alpha_t}}x_0 + \sqrt{1-\overline{\alpha_t}}\epsilon, \quad \epsilon \sim \mathcal{N}(0, I) \tag{3}$$

It is worth noting that the noise $\epsilon$ serves as the training label that the network predicts. The noisy signal $x_t$ and time embedding were fed into the model. Through feature extraction, the model ultimately predicted noise $\epsilon_\theta(x_t, t)$ as its output.

### 2.5. Enhanced Loss Computation

We employed an enhanced weighted loss function to evaluate discrepancy between the predicted noise $\epsilon_\theta(x_t, t)$ and its ground truth $\epsilon$. The Smooth L1 loss was applied to noise residual $r_i = \epsilon_i - \epsilon_{\theta i}$ as the base loss, where $i$ indexes the signal sampling points:

$$L_{base}(r_i) = \begin{cases} 0.5 r_i^2, & \text{if } |r_i| < 1 \\ |r_i| - 0.5, & \text{if } |r_i| \geq 1 \end{cases} \tag{4}$$

This loss balances sensitivity to small errors with robustness to outliers. To jointly enhance amplitude restoration and state transition capture, we introduced a dynamic weighting mechanism, consisting of an amplitude weight and an edge weight. The amplitude weight emphasizes regions with large noise residuals, encouraging the model to suppress strong disturbances:

$$W_{amp}^{(i)} = 1 + \lambda_{amp} |\epsilon_i| \tag{5}$$

where, parameter $\lambda_{amp}$ controls the sensitivity of the loss function to amplitude. The edge weight captures location and magnitude of transition edges, based on the first- and second-order differences ($\nabla x$ and $\nabla^2 x$) of the noise-free signal $x_0$. These differences are smoothed using a 1×3 convolution kernel $K = [1/3, 1/3, 1/3]$, expanding sparse transition points into a localized penalty band:

$$W_{edge}^{(i)} = 1 + \lambda_{edge} [\text{Conv1d}(\nabla x + 0.5 \nabla^2 x, K)]_i \tag{6}$$

where, parameter $\lambda_{edge}$ adjusts the emphasis on the detection errors of transition points. The final training objective is defined as a weighted average over all signal trace:

$$L_{total} = \frac{1}{N} \sum_{i=1}^{N} \left( W_{amp}^{(i)} \cdot W_{edge}^{(i)} \cdot L_{base}^{(i)} \right) \tag{7}$$

where, $N$ is total number of signal sampling points, $W^{(i)}_{amp}$ and $W^{(i)}_{edge}$ are the amplitude and edge weight of the $i$-th sampling point, and $L^{(i)}_{base}$ is base loss at the $i$-th point. Dedicated hyperparameter tuning was performed on the amplitude and edge weighting factors, $\lambda_{amp}$ and $\lambda_{edge}$, (see Supporting Note 4 in the SI).

## 2.6. Performance evaluation

To quantitatively evaluate the denoising performance of SSDM, we developed a comprehensive assessment framework that considers both signal amplitude recovery and accurate identification of physically meaningful state transitions, as illustrated in Figure 1 (c). Amplitude accuracy was measured using the mean squared error (*MSE*) between the denoised trajectory $\hat{x}$ and the signal ground truth $x_{gt}$.

$$MSE = \frac{1}{N} \sum_{t=1}^{N} (\hat{x}_t - x_{gt,t})^2 \tag{8}$$

Transition-point detection accuracy was evaluated using a tolerance-based matching criterion. A predicted transition point $t_{pred}$ is considered correct if it satisfied $|t_{gt} - t_{pred}| \leq \delta$, with the tolerance set to be $\delta = 2$. Based on the matching results, we computed the *F1* score, which balances precision and recall:

$$F1 = 2 \cdot \frac{\text{Precision} \cdot \text{Recall}}{\text{Precision} + \text{Recall}} \tag{9}$$

with Precision $= \frac{TP}{TP+FP}$ and Recall $= \frac{TP}{TP+FN}$, where, *TP*, *FP*, and *FN* denote true positives, false positives, and false negatives, respectively. To identify the transition points, we employed amplitudes thresholds, which were predefined based on the number of states: $A_{th,2} = \{0.5\}$ for 2-state signals, $A_{th,3} = \{0.25, 0.75\}$ for 3-state signals, and $A_{th,4} = \{0.165, 0.5, 0.83\}$ for 4-state signals. Each time point was assigned to a state by threshold comparison. To provide a unified performance measure, we defined a composite score combining amplitude and transition accuracy:

$$Score = \log\left(\frac{F1}{MSE}\right) \tag{10}$$

All models were implemented using PyTorch 2.7.1 with CUDA 11.8 (cu118) and trained on a single NVIDIA L40S GPU with 48 GB memory. Training on the full dataset of 10,800 signals required approximately 23 hours. Our implementation is partially modified from the open-source PyTorch code provided in the GitHub repository "*yet-another-pytorch-tutorial-v2*" by Sungjoon Choi[49].

## 3. Results and discussion

### 3.1. Performance on the generated dataset

SSDM was successfully trained and achieved good performance on the test dataset with *MSE* = 0.0041, *F1* = 0.96, and *Score* = 8.31. As a typical example with SNR = 3 shown in Figure 2 (a), the denoised signal matches the ground truth perfectly. It indicates the model can effectively identify the transition points and the state levels. The performance of SSDM depends on signal quality, i.e., SNR, as shown in Figure 2 (b). With the increase of SNR, the impact of noise decreases, and, thus, the model gets a higher *Score* in general for all kinds of signals, *i.e.*, 2-, 3-, and 4-state. The *Score* reaches a plateau at SNR = 3, indicating that further increases in the SNR have a limited contribution to improvement of denoising performance. Model performance only shows a slight variation on different kinds of signals, reflecting its adaptability to complexity of the signals. At lower SNRs (≤ 1), signals with fewer states exhibit lower *Score*, as noise fluctuations are more likely to be misidentified as state transitions.

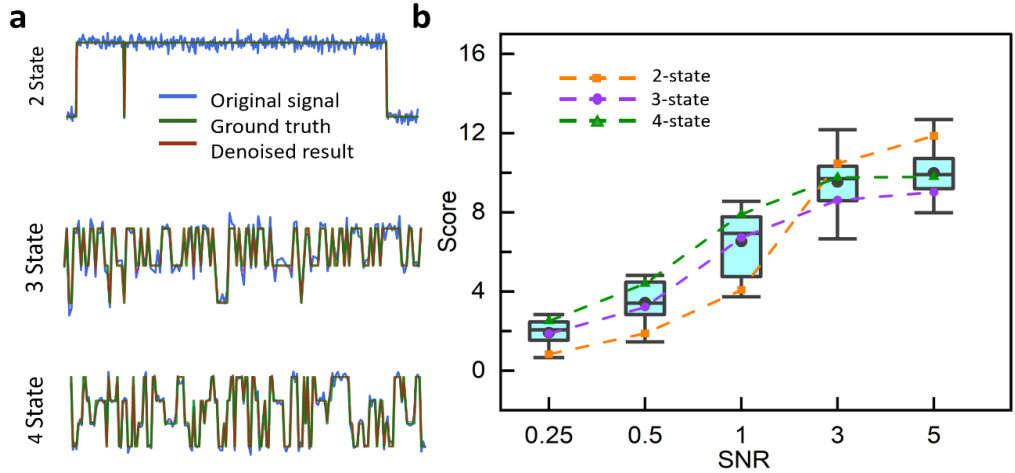

Figure 2. Denoising performance of SSDM for signals with different numbers of states and SNRs. (a) Typical examples of denoised signals with 2-, 3-, and 4-state, respectively. The blue curves are original noisy signals with SNR = 3, green curve is the ground truth, and the red curves show the designed results. (b) The *Score* of the denoising performance of SSDM for signal with different SNRs. Box chart shows the results from the whole dataset. The box spans the 25$^{th}$ to 75$^{th}$ percentile, with the central line indicating the median. Mean values are marked as black dots. Whiskers extend from the 5$^{th}$ to the 95$^{th}$ percentile. The dot-on lines represent the results of the 2-, 3-, and 4-state signals from the dataset, respectively.

We further analyze *F1* and *MSE* of the denoised signals with different dynamic properties, as shown in Figure 3. It was observed that when SNR > 1, *F1* approaches to one with minimal variation for all 2-, 3-, and 4-state signals with different average state durations ($\tau_{a,b,c,d}$, the average time of the signal staying at state a, b, c, and d), as shown in Figure 3 (b, f, j). It indicates that SSDM performs well in both accuracy and stability for identifying transition points under higher SNR conditions. When SNR < 1, *F1* shows a general trend of a slight decline as the duration of other states increases, especially for 3- and 4-state signals (see Figure 3 (a, e, i)). It suggests that under lower SNR conditions, SSDM performs better in the identification of transition points for signals with more frequent state transitions (smaller $\tau$). It can be attribute to the fact that the signals with larger $\tau$ have less transition events in 1000-point trace, which offers less learning examples for the model to acquire the ability of identification of transition points.

In contrast, *MSE* demonstrates more complicated trends by varying state duration, $\tau$. For 2-state signals, *MSE* decreases in general as $\tau_{a,b}$ increases for all SNRs, as shown in Figure 3 (c, d). If a state has a long duration, the signal possesses relatively sparse transitions, which eases the state level prediction and decreases *MSE*. For 3- and 4-state signals (Figure 3 (g, h, k, i)), *MSE* is no longer determined solely by state duration, $\tau_{a,b,c,d}$, and more factors are involved, such as relative proportions of the durations among different states and state transition rates. *MSE* exhibits non-monotonic trends, with a strong dependence on SNR.

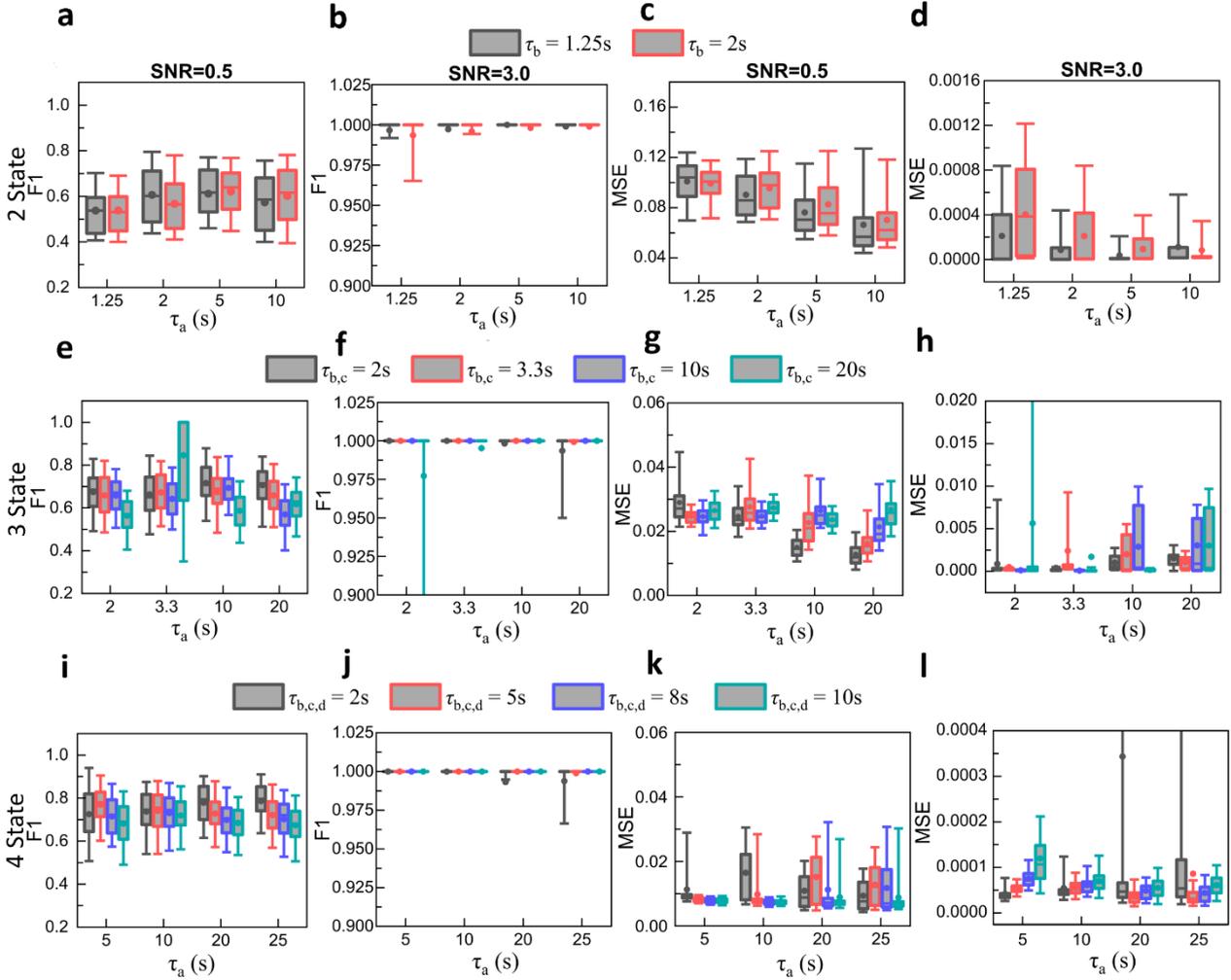

Figure 3. Performance of SSDM for signals with different state durations and SNRs. (a–d) *F1* and *MSE* for 2-state signals at SNR = 0.5 and 3, respectively. (e–h) *F1* and *MSE* for 3-state signals at SNR = 0.5 and 3, respectively. (i–l) *F1* and *MSE* for 4-state signals at SNR = 0.5 and 3, respectively. The box spans the 25th to 75th percentiles, with the central line indicating the median. Mean values are marked as dots, and whiskers extend from the 5th to the 95th percentile.

### 3.2 Comparison of generalized and specialized models

To further evaluate the model performance for signals with a known number of states, we trained specialized models for 2-, 3-, and 4-state, respectively, by using independently generated training and test sets for each model. The data size, ranges of corresponding parameters, and ratio among parameter combinations are the same as those used for the training and test of the generalized model in previous section. Figure 4 (a–c) presents representative examples comparing ground truth signals with denoising results obtained from the specialized and generalized models for the same signal segments with 2, 3, and 4 states at SNR = 1, respectively. While both models effectively denoise the signals, the specialized models achieve superior performance in preserving fine details. In particular, the generalized model occasionally misrecognizes noise fluctuations as state transitions, most notably in the 2-state case, whereas this issue is largely mitigated by the specialized models.

Overall, as the SNR increases from 0.25 to 5, the *Score* of all models show a significant increase. In general, all specialized models show a better performance than the generalized model, especially for large SNR signals, as compared in Figure 4 (d). In addition, the performance of specialized models has less variations indicating more stable outcomes. It suggests that if only a tiny bit more information, the number of states, is known, it can significantly help enhance the denoising performance of SSDM, since the task of processing certain kinds of signals is easier than handling all possible variations, in agreement with our intuitive understanding.

Under lower SNR conditions, particularly when SNR < 1(Figure 4 (d)), the 4-state model achieves the highest *Score*, whereas the 2-state model performs the worst. However, as the SNR increases beyond 1, the performance trend reverses. The 2-state model exhibits a clear advantage and outperforms other models. Given that the *Score* is determined by both *F1* and *MSE*, we further examined these two metrics separately. The results indicate that when SNR ≥ 1, *F1* and *SME* among different models are similar (Figure 4 (e, f)), and the generalized model shows a slightly worse performance, i.e., a lower *F1* and a higher *SME*. When SNR < 1, the 2-state model shows higher *MSE* (Figure 4(f)), indicating that its signal reconstruction accuracy is more heavily impacted by noise. Under low SNRs, noise causes rapid fluctuations in the signal. If a model has few states, it is difficult to distinguish these fluctuations from the real state transitions. However, models with more states can incorporate short-duration states to describe these fluctuations, enabling more accurate signal reconstruction.

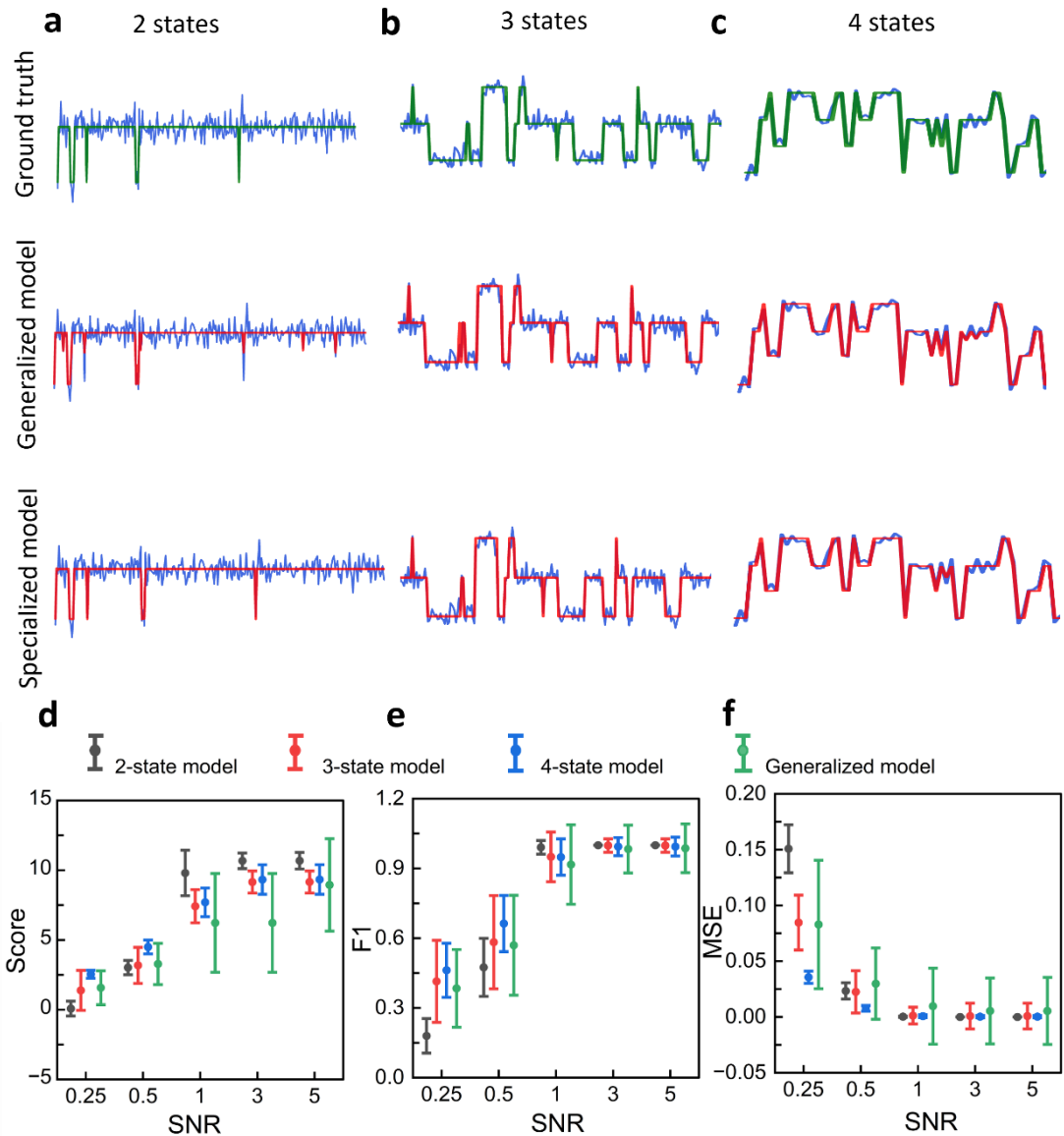

Figure 4. Denoising performance comparison between the specialized and generalized SSDMs. (a-c) Typical examples of ground truth and denoising results obtained from the specialized and generalized models for the same signal segments with 2, 3, and 4 states at SNR = 1. The blue curves are original noisy signals, green curves are the ground truth, and red curves show the denoised results. (d-f) The *Score*, *F1*, and *MSE* of denoising signals processed by generalized models and specialized models, respectively. The error bars denote the standard deviation, mean values are marked as dots.

## 3.3. Comparison with other methods

To better assess the performance, we benchmark the SSDM with low-pass filters and HMM by evaluating the results with the same metrics. We implemented a 4-order low pass filtering to perform filtering operations on signals. In terms of parameter selection, we performed filtering on the complete test dataset with a set of candidate cutoff frequencies, assessed *F1*, *MSE*, and the *Score* for each cutoff frequency, and selected one with the highest *Score* as the final optimized filter. For detailed implementation, refer to Supporting Note 5 in the SI. A typical example of the denoised signals from the low-pass filter with the optimized cutoff frequency is shown in Figure 5 (a) and the *Score* on the entire test dataset with different SNRs is shown in Figure 5 (b). It shows a much worse performance compared to that of the SSDM. The filtering results highly rely on the selection of the cutoff frequency, since the frequency components of the signal and noise are highly overlapped in spectrum (see Figure S2 in the SI), which is an inherent physical limit of low-pass filters and cannot be avoided.

HMM focuses on recovering hidden state sequences and it decodes the most probable hidden state at each time point rather than directly denoises signals. By assigning a state for each time point, HMM outputs the mean value of the corresponding state and the output sequence looks like a denoised signal. Although it is not a model specifically designed for denoising, it can still perform the noise reduction in a certain degree. In HMMs, the number of hidden states is a critical parameter. Usually, the true number of hidden states in stepwise signals is usually unknown. To find the optimal number of states, we select the state count by iterating through possible state numbers with Bayesian Information Criterion (BIC). For detailed implementation, refer to Supporting Note 6 in the SI. The same example of noisy signal treated by HMM is shown in Figure 5 (a). We can see that the denoised signal totally misses one state. Similarly, we employed the *Score* to evaluate the performance, as shown in Figure 5 (b). HMM relies on the emission probability of each state to determine which state the observation belongs to. Under conditions with lower SNR, the distribution of different states strongly overlap, which causes a significant difficulty in state recognition, and induces numerous false transitions and state missing. These result in a poor *F1* and *MSE*, so does the overall *Score*. At higher SNR, signals become clear, enabling HMM to identify state sequences more readily. However, this extreme clarity of signal introduces overfitting issues. Thus, when SNR $\geq$ 1, the variance in HMM *Score* increases significantly. In summary, SSDM achieves higher *Score* compared to low-pass filter and HMM. Although the *Score* of HMM is nearly as good as SSDM, it suffers much larger variations especially for SNR $\geq$ 1 and, thus, may deliver less stable denoising results.

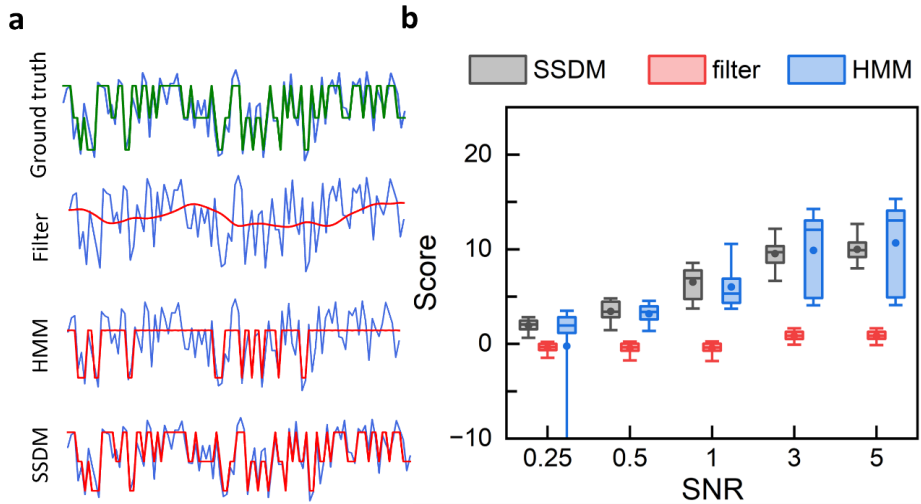

Figure 5. Benchmark of the denoising performance of SSDM with low-pass filter and HMM. (a) Typical example of a 3-state signal segment with SNR = 0.5. Ground truth (green) and denoising signals (red) processed by Low-pass filter, HMM, and SSDM are shown, respectively. (b) The *Score* of denoising signals with different SNRs processed by low-pass filter, HMM, and SSDM. The box spans the $25^{th}$ to $75^{th}$ percentile, with the central line indicating the median. Mean values are marked as dots, and whiskers extend from the $5^{th}$ to the $95^{th}$ percentile.

### 3.4. Validation by experimental data

To evaluate the effectiveness and generalization capacity of SSDM, we applied the trained SSDM to real-world experimental data from sm-FRET and nanopore λ-DNA translocation. As a typical example shown in Figure 6 (a), after SSDM denoising, the FRET efficiency trace shows a clear two-state structure. SSDM significantly reduces its background noise while honestly preserves the transition positions present in the original signal. Afterwards, each data point of the denoised signals is assigned to state 1 or 2 by referring to an amplitude threshold, which is simply selected as the midpoint value between the two state levels. Furthermore, the kinetic parameter of the two-state transition can be extracted. The dwell times of each state coincide with exponential distributions (see Figure 3 in the SI), indicating Poisson processes of the state transition in agreement with physical understanding. The corresponding kinetic rate constants $k_{12}$ and $k_{21}$, i.e., the average transition frequency from state 1 to 2 or state 2 to 1, are extracted from the distributions.

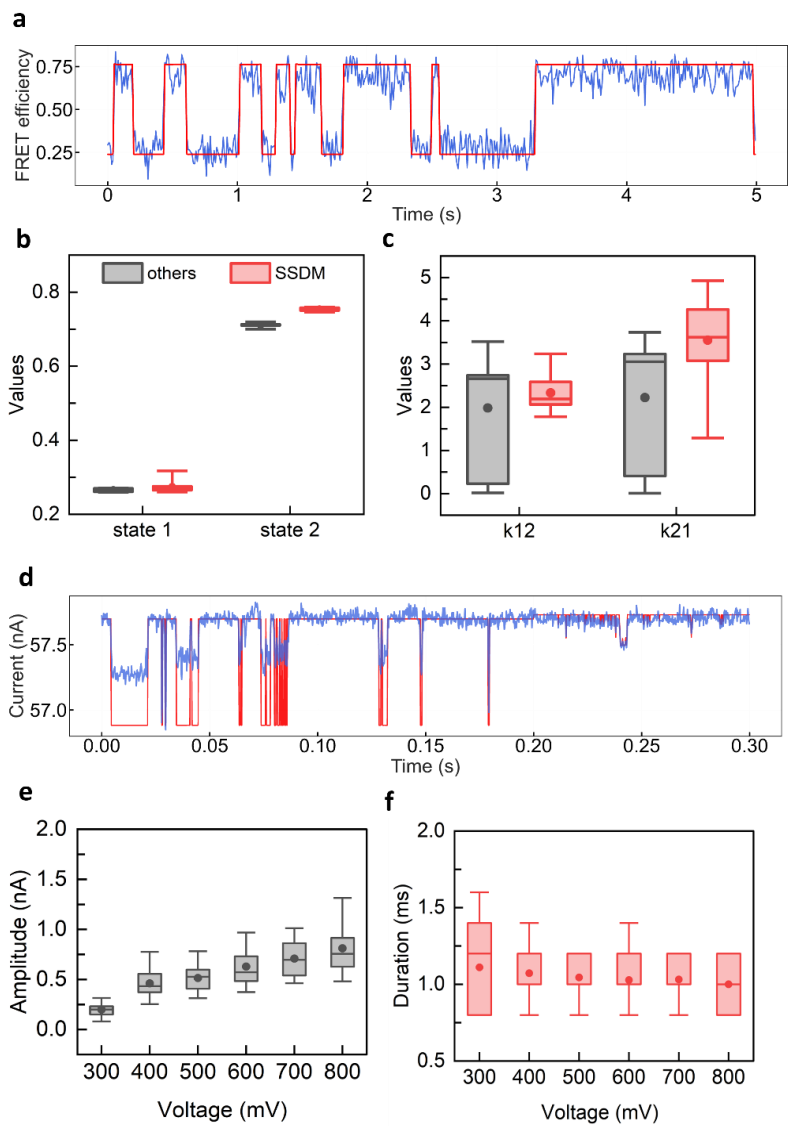

Figure 6. Signal processing results of the experimental data by SSDM. (a) Typical example of sm-FRET efficiency trajectory denoised using SSDM. The blue curve is original noisy signals, the red curve shows the designed results. (b, c) Comparison of extracted state levels and kinetic parameters of smFRET state transitions obtained using SSDM and other analysis methods, respectively. (d) Typical example of nanopore λ-DNA translocation signal processed by SSDM. The blue curve is original noisy signals, and the red curve shows the designed results. (e, f) Amplitude and duration of the λ-DNA translocation events at different bias voltages. All box plots in the figure follow the same format: the box spans the 25[th] to 75[th] percentile, with the central line indicating the median. Mean values are marked as dots, and whiskers extend from the 5[th] to the 95[th] percentile.

We compared the extracted state levels and kinetic rate constants with those from other 14 different tools on the same dataset in ref[10], as shown in Figure 6 (b, c). It can be seen that the state levels estimated from SSDM denoised signals locate at ~ 0.25 and ~ 0.70, which match well with those obtained from the other tools. The kinetic rate constant $k_{12}$ is close to the mean value obtained from other tools, while $k_{21}$ is slightly higher than that from other tools. This deviation may come from the error of state assignment by using a simple threshold, and the results are sensitive to the threshold position. Overall, SSDM can effectively restore the bistable structure from FRET signals with good stability and reliability.

Furthermore, we employed SSDM to denoise and analyze signals of λ-DNA nanopore translocation at different bias voltages ranging from 300 mV to 800 mV. As a typical example shown in Figure 6 (d), after SSDM denoising, we can clearly see the translocation events without background noise fluctuations. The translocation events are extracted by applying an amplitude threshold. The amplitude and duration of each translation event are further measured. We can clearly see from Figure 6 (e, f) that the amplitude monotonically increases with increasing voltage, while the duration decreases as voltage increases. This trend complies with the physical understanding: a higher voltage generates a larger ionic current so does the translocation spikes; a higher voltage drives the translocation of DNA faster through the nanopore, indicating a short duration. In addition, the results are consistent with those from other algorithms[48,50]. Overall, the results support the validity of SSDM, which can efficiently remove the noise while preserving the signal features.

### 3.5. Discussion

SSDM demonstrates superior denoising performance on stepwise signals, particularly in its ability to preserve sharp state transitions while effectively suppressing noise. To further adapt and optimize the model for diverse real-world applications, several practical considerations should be taken into account.

The length of input signal segment directly affects the inference time in the reverse diffusion process. In current implementation, SSDM operates on fixed 1000-point signal segments. However, this length can be adjusted according to signal properties in different applications, allowing a balance among segment length, denoising performance, and computation efficiency.

Furthermore, the standard DDPM framework assumes Gaussian white noise as the default noise model. As a results, during training, the model naturally develops a preference for removing random white noise. For signals with different statistical characteristics of noise, model performance can be further improved by generating specified training dataets that include corresponding noise.

For more challenging scenarios, such as signals with extremely low SNR, a large number of states, and various state levels, additional optimization of SSDM hyperparameter is necessary. The trade-off between accurate detection of transition points and fitting state levels can be controlled by tuning the loss function. Training efficiency may also be improved by optimizing the evaluation metrics.

For applications demand Real-time data processing, inference latency becomes a critical factor. In such cases, advanced versions of diffusion-based models may be adopted, such as Denoising Diffusion Implicit Models and other accelerated variants[51]. Moreover, when deployment on embedded low-power hardware platforms is required, a careful balance between model complexity and lightweight design should be considered.

### 4. Conclusion

In this work, we propose SSDM, a diffusion-based framework for denoising stepwise signals. The model adopts an iterative denoising strategy that combines a U-Net architecture with attention mechanisms to effectively capture both local features and long-range dependencies. To quantitatively assess denoising performance for stepwise signals, we introduce a set of task-specific metrics that jointly evaluate state-level reconstruction accuracy and state-transition identification. Through tailored loss functions and evaluation strategies, SSDM achieves improved denoising accuracy while enhancing the reliability of state and transition detection. Benchmark results demonstrate that SSDM substantially reduces noise from stepwise signals even under severe noise conditions with SNR < 1. Furthermore, specialized models trained for 2-, 3-, and 4-state signals outperform a generalized model, indicating that even limited prior knowledge of the number of states can significantly enhance performance. Compared with traditional filtering approaches and hidden Markov models, SSDM achieves superior performance in terms of the overall denoising *Score*. The effectiveness of SSDM is further validated using real experimental data from sm-FRET and nanopore DNA translocation measurements. The denoised results are consistent with those obtained using other established methods and align with physical expectations, demonstrating the strong generalization capability of SSDM. Overall, SSDM provides a robust and flexible framework for stepwise signal denoising and shows significant promise for addressing complex signal processing challenges across a wide range of applications.

**Acknowledgement**

The authors thank Dr. Dario Dematties for valuable technical advice on diffusion model. The authors thank Sungjoon Choi for making the open-source implementation publicly available. The authors thank the National Academic Infrastructure for Supercomputing in Sweden (NAISS) for providing the computational resources through the Alvis cluster at Chalmers Centre for Scientific Science and Engineering (C3SE, NAISS 2025/22-1024) and the Pelle cluster at the Uppsala Multidisciplinary Center for Advanced Computational Science (UPPMAX 2025/2-328). This work was partially supported by Vetenskapsrådet (2025-05333).

**Notes**

The authors declare no competing financial interest. The generated data sets for all the combinations of the parameters are available at https://doi.org/10.5281/zenodo.18432190. The code for all SSDMs used in this work is available at https://doi.org/10.5281/zenodo.18417193.

**Supplementary Information**

# Denoise Stepwise Signals by Diffusion Model Based Approach

Xingdi Tong[1], Chenyu Wen[2*]

1. Department of Civil and Industrial Engineering, Ångströmlaboratoriet, Uppsala University, Lägerhyddsvägen 1, 75237, Uppsala, Sweden

2. Division of Solid-State Electronics, Department of Electrical Engineering, Ångströmlaboratoriet, Uppsala University, Lägerhyddsvägen 1, 75237, Uppsala, Sweden

* Corresponding author: Chenyu Wen: chenyu.wen@angstrom.uu.se

**Supporting Note 1. Network architecture**

We designed a one-dimensional (1D) convolutional neural network (CNN) based on the U-Net architecture to predict noise $\epsilon_\theta(x_t, t)$. This network adopts an encoder-decoder structure and fuses multi-scale features through skip connections. The entire network consists of a four-layer downsampling encoder and a four-layer upsampling decoder, forming a symmetrical structure with a depth of four layers. The input of model is a 1D noisy stepwise signals $x_t$ with shape 1×1000. The embedding vector at diffusion time step $t$ is also fed into the network, enabling the model to perform noise prediction at different noise levels.

In the encoder process, to enhance model capacity, the size of the channels is scaled up from 192 to 384, 768, and 1536. In each layer, the network performs a downsampling on signal, thereby increasing the temporal scope captured by subsequent convolutions. In the final bottleneck layer, the network learns the overall trend of signal. Then, the decoder performs upsampling layer by layer with a symmetric structure as the encoder. Along the flow, it merges the corresponding encoder features into the decoding path through skip connections at each layer. This approach prevents detail loss caused by downsampling, enabling the network to simultaneously learn local fine structures and global general patterns.

The main body of the encoder and decoder consists of multiple ResBlocks. In each ResBlock, local temporal features are first extracted by a 1D convolution with a kernel size of three and padding of one. Then, the SiLU activation function is applied to introduce a nonlinear transformation. Feature maps are subjected to adaptive group normalization, which modulates the normalized features using scale and shift parameters derived from temporal step embeddings. This enables adaptive adjustment of feature representations across different diffusion time steps.

Convolutional structures can efficiently capture local patterns but have limited capacity to model global dependencies across time steps. To address this issue, we adopted a full-scale attention integration strategy, considering the lower computational complexity of 1D signals. In each layer of both the encoder and decoder, an Attention Block was connected after the ResBlock.

To improve the stability of gradient propagation in deep networks, input feature maps $x \in \mathbb{R}^{C \times L}$, where $C$ is the number of channels, and $L$ is the length of feature, first proceed through a group normalization layer. The channel dimension is divided into 16 groups performing independent normalization within each group. The normalized features are expanded to $3C$ channels via a convolutional layer and split into query, $Q$, key, $K$, and value, $V$. To learn information from multiple perspectives, features are divided into four attention heads, each independently processing its sub-features. For each head, compute the dot product between $Q$ and $K^T$ to generate correlation matrix $QK^T$. After Softmax normalization, this matrix serves as weights for aggregating $V$, enabling each time step to incorporate information from the entire sequence. The aggregated global features proceed through an output projection layer for channel integration. In the output layer, initial weights and biases are set to zero, causing the attention module to behave as an identity mapping during early stage of training. This strategy enables the model to gradually learn and superimpose global residual corrections while preserving local convolutional features.

**Supporting Note 2. Data generation**

We simulated stepwise signals by a 1D Markov chain. By allocating a fixed amplitude to each state, the segmented constant values of the signal are obtained. State transitions are determined by the transition rate matrix $M$. The diagonal elements $m_{ii}$ of the matrix govern the distribution of state residence times, leading to an average dwell time $\tau_i$ determined by $m_{ii}$, where $\tau_i = 1/|m_{ii}|$. The off-diagonal element $m_{ij}$ represents the rate at which the system transitions from state $i$ to state $j$. Subsequently, Gaussian pink noise of varying intensities is added to generate the corresponding noisy signals.

To construct noisy stepwise signals, we generate zero-mean Gaussian noise with a controllably designed different RMS level, corresponding to different target SNRs. SNR is defined as the ratio of the spike minimum difference among levels, as amplitude to the peak-to-peak value of the background noise. For Gaussian noise, the peak-to-peak value is typically estimated as six times the root-mean-square value noise. By adjusting different SNR values, we can obtain noise of varying levels. After noise is generated, it will be added to the pre-generated stepwise signals to form noisy signals.

We constructed a balanced dataset comprising three types of single-molecule stepwise signals: 2-state, 3-state, and 4-state. All signals were generated at a sampling rate of 1 Hz, with each fixed at 1000 time points. A total of 10,800 signals samples were generated for training and optimizing. To keep the data simple, state value of all signals is defined using a unified normalization method. The state levels uniformly distributed in the normalized interval [0, 1]. In other words, the 2-state signal uses state values of possesses two levels located at 0 and 1. The 3-state signal has three levels at 0, 0.5, and 1, and the 4-state signal has four levels at 0, 0.33, 0.66, and 1.

To achieve a gradual progression in dynamic complexity and recognition difficulty across different signal types while ensuring comprehensive coverage of the parameter space, we designed tailored generation logic for signal type. For the 2-state signals, eight representative combinations of transition rates were selected, as shown in Table S1. For the 3-state signals, the transition rate matrix contains more independent parameters. First, we set the escape rates of the three states, $m_{ii}$, $i = 1, 2,$ and 3, to be identical, which is divided into six rate categories, ranging from extremely slow (0.01), slow (0.05), moderately slow (0.1), moderate (0.3), fairly fast (0.5), to fast (0.8). Subsequently, within each rate group, two representative internal transition modes were selected. By combining six escape rate types with two internal modes, 12 typical 3-state transition rate matrices were ultimately obtained, as shown in Table S1. In the 4-state signals, the escape rate of one reference state was fixed as an anchor point, while the escape rates of the other three states were varied around it. 16 representatives transition rate matrices were generated, as shown in Table S1. Each matrix generated 100 samples, and covering SNR = 1, SNR = 3, and SNR = 5 types. Therefore, there are 2400, 3600, and 4,800 samples for 2-, 3-, and 4-state signals, respectively.

Table S1. Signal Parameter for train dataset

| State Number | Transition rate matrix (M) | State Number | Transition rate matrix (M) | State Number | Transition rate matrix (M) |
|---|---|---|---|---|---|
| 2 | $\begin{bmatrix} -0.01 & 0.01 \\ 0.01 & -0.01 \end{bmatrix}$ | 3 | $\begin{bmatrix} -0.01 & 0.005 & 0.005 \\ 0.005 & -0.01 & 0.005 \\ 0.005 & 0.005 & -0.01 \end{bmatrix}$ | 4 | $\begin{bmatrix} -0.05 & 0.01 & 0.02 & 0.02 \\ 0.015 & -0.05 & 0.015 & 0.02 \\ 0.012 & 0.018 & -0.05 & 0.02 \\ 0.019 & 0.011 & 0.02 & -0.05 \end{bmatrix}$ |
| | | | $\begin{bmatrix} -0.01 & 0.002 & 0.008 \\ 0.002 & -0.01 & 0.008 \\ 0.002 & 0.008 & -0.01 \end{bmatrix}$ | | $\begin{bmatrix} -0.05 & 0.01 & 0.02 & 0.02 \\ 0.15 & -0.35 & 0.05 & 0.15 \\ 0.012 & 0.018 & -0.05 & 0.02 \\ 0.019 & 0.011 & 0.02 & -0.05 \end{bmatrix}$ |
| | $\begin{bmatrix} -0.01 & 0.01 \\ 0.02 & -0.02 \end{bmatrix}$ | | $\begin{bmatrix} -0.05 & 0.02 & 0.03 \\ 0.02 & -0.05 & 0.03 \\ 0.02 & 0.03 & -0.05 \end{bmatrix}$ | | $\begin{bmatrix} -0.05 & 0.01 & 0.02 & 0.02 \\ 0.005 & -0.05 & 0.025 & 0.02 \\ 0.12 & 0.18 & -0.5 & 0.2 \\ 0.019 & 0.011 & 0.02 & -0.05 \end{bmatrix}$ |

| | | | |
|---|---|---|---|
| | | | $\begin{bmatrix} -0.05 & 0.01 & 0.02 & 0.02 \\ 0.005 & -0.05 & 0.025 & 0.02 \\ 0.012 & 0.018 & -0.05 & 0.02 \\ 0.06 & 0.01 & 0.03 & -0.1 \end{bmatrix}$ |
| $\begin{bmatrix} -0.01 & 0.01 \\ 0.05 & -0.05 \end{bmatrix}$ | | $\begin{bmatrix} -0.05 & 0.04 & 0.01 \\ 0.04 & -0.05 & 0.01 \\ 0.04 & 0.01 & -0.05 \end{bmatrix}$ | $\begin{bmatrix} -0.1 & 0.03 & 0.04 & 0.03 \\ 0.05 & -0.1 & 0.02 & 0.03 \\ 0.03 & 0.03 & -0.1 & 0.04 \\ 0.05 & 0.02 & 0.03 & -0.1 \end{bmatrix}$ |
| | | $\begin{bmatrix} -0.1 & 0.05 & 0.05 \\ 0.05 & -0.1 & 0.05 \\ 0.05 & 0.05 & -0.1 \end{bmatrix}$ | $\begin{bmatrix} -0.1 & 0.03 & 0.04 & 0.03 \\ 0.1 & -0.3 & 0.1 & 0.1 \\ 0.1 & 0.2 & -0.5 & 0.2 \\ 0.019 & 0.011 & 0.02 & -0.05 \end{bmatrix}$ |
| $\begin{bmatrix} -0.01 & 0.01 \\ 0.1 & -0.1 \end{bmatrix}$ | | $\begin{bmatrix} -0.1 & 0.01 & 0.09 \\ 0.01 & -0.1 & 0.09 \\ 0.01 & 0.09 & -0.1 \end{bmatrix}$ | $\begin{bmatrix} -0.1 & 0.03 & 0.04 & 0.03 \\ 20 & -50 & 20 & 10 \\ 0.1 & 0.2 & -0.5 & 0.2 \\ 0.019 & 0.011 & 0.02 & -0.05 \end{bmatrix}$ |
| | | | $\begin{bmatrix} -0.1 & 0.03 & 0.04 & 0.03 \\ 20 & -50 & 20 & 10 \\ 30 & 20 & -60 & 10 \\ 0.019 & 0.011 & 0.02 & -0.05 \end{bmatrix}$ |
| $\begin{bmatrix} -0.01 & 0.01 \\ 0.5 & -0.5 \end{bmatrix}$ | | $\begin{bmatrix} -0.3 & 0.1 & 0.2 \\ 0.1 & -0.3 & 0.2 \\ 0.1 & 0.2 & -0.3 \end{bmatrix}$ | $\begin{bmatrix} -0.3 & 0.07 & 0.11 & 0.12 \\ 0.1 & -0.3 & 0.1 & 0.1 \\ 0.08 & 0.12 & -0.3 & 0.1 \\ 0.09 & 0.08 & 0.13 & -0.3 \end{bmatrix}$ |
| | | $\begin{bmatrix} -0.3 & 0.05 & 0.25 \\ 0.05 & -0.3 & 0.25 \\ 0.05 & 0.25 & -0.3 \end{bmatrix}$ | $\begin{bmatrix} -0.3 & 0.07 & 0.11 & 0.12 \\ 0.07 & -0.3 & 0.11 & 0.12 \\ 0.05 & 0.05 & -0.3 & 0.2 \\ 0.8 & 0.2 & 0.5 & -0.15 \end{bmatrix}$ |
| $\begin{bmatrix} -0.05 & 0.05 \\ 0.01 & -0.01 \end{bmatrix}$ | | $\begin{bmatrix} -0.5 & 0.2 & 0.3 \\ 0.2 & -0.5 & 0.3 \\ 0.2 & 0.3 & -0.5 \end{bmatrix}$ | $\begin{bmatrix} -0.3 & 0.07 & 0.11 & 0.12 \\ 0.2 & -0.8 & 0.3 & 0.3 \\ 0.07 & 0.11 & -0.3 & 0.12 \\ 0.07 & 0.12 & 0.11 & -0.3 \end{bmatrix}$ |
| | | | $\begin{bmatrix} -0.3 & 0.07 & 0.11 & 0.12 \\ 0.07 & -0.3 & 0.11 & 0.12 \\ 0.5 & 0.05 & -0.1 & 0.25 \\ 0.07 & 0.12 & 0.11 & -0.3 \end{bmatrix}$ |
| $\begin{bmatrix} -0.2 & 0.2 \\ 0.01 & -0.01 \end{bmatrix}$ | | $\begin{bmatrix} -0.5 & 0.1 & 0.4 \\ 0.1 & -0.5 & 0.4 \\ 0.1 & 0.4 & -0.5 \end{bmatrix}$ | $\begin{bmatrix} -0.5 & 0.1 & 0.2 & 0.2 \\ 0.15 & -0.5 & 0.15 & 0.2 \\ 0.12 & 0.18 & -0.5 & 0.2 \\ 0.19 & 0.11 & 0.2 & -0.5 \end{bmatrix}$ |
| | | $\begin{bmatrix} -0.8 & 0.4 & 0.4 \\ 0.4 & -0.8 & 0.4 \\ 0.4 & 0.4 & -0.8 \end{bmatrix}$ | $\begin{bmatrix} -0.5 & 0.1 & 0.2 & 0.2 \\ 0.015 & -0.1 & 0.015 & 0.07 \\ 0.12 & 0.18 & -0.5 & 0.2 \\ 0.2 & 0.3 & 0.5 & -1 \end{bmatrix}$ |
| $\begin{bmatrix} -0.5 & 0.5 \\ 0.01 & -0.01 \end{bmatrix}$ | | $\begin{bmatrix} -0.8 & 0.1 & 0.7 \\ 0.1 & -0.8 & 0.7 \\ 0.1 & 0.7 & -0.8 \end{bmatrix}$ | $\begin{bmatrix} -0.5 & 0.1 & 0.2 & 0.2 \\ 0.015 & -0.1 & 0.015 & 0.07 \\ 0.12 & 0.18 & -0.5 & 0.2 \\ 0.19 & 0.11 & 0.2 & -0.5 \end{bmatrix}$ |
| | | | $\begin{bmatrix} -10 & 3 & 4 & 3 \\ 0.2 & -0.5 & 0.2 & 0.1 \\ 0.3 & 0.2 & -0.6 & 0.1 \\ 19 & 11 & 20 & -50 \end{bmatrix}$ |

The test dataset was generated independently from the training dataset to evaluate the generalization ability and robustness of SSDM. Compared with the training data, the test dataset mainly consists of signals with more rapid state transitions, corresponding to higher escape rates, as shown in the Table S2. Within the whole test dataset, a total of 40 distinct parameter combinations were included, comprising 8 for 2-state signals, 16 for 3-state signals, and 16 for 4-state signals. Each combination contains 20 samples for each SNR level (SNR = 0.25, 0.5, 1, 3, and 5), resulting in 4,000 samples per SNR level and 20,000 samples in total.

Table S2. Signal Parameter for test dataset

| State Number | Transition rate matrix (M) | State Number | Transition rate matrix (M) | State Number | Transition rate matrix (M) |
|---|---|---|---|---|---|
| 2 | $\begin{bmatrix} -0.1 & 0.1 \\ 0.5 & -0.5 \end{bmatrix}$ | 3 | $\begin{bmatrix} -0.1 & 0.05 & 0.05 \\ 0.05 & -0.1 & 0.05 \\ 0.05 & 0.05 & -0.1 \end{bmatrix}$ | 4 | $\begin{bmatrix} -0.1 & 0.03 & 0.04 & 0.03 \\ 0.03 & -0.1 & 0.04 & 0.03 \\ 0.03 & 0.04 & -0.1 & 0.03 \\ 0.03 & 0.04 & 0.03 & -0.1 \end{bmatrix}$ |
| | | | $\begin{bmatrix} -0.1 & 0.05 & 0.05 \\ 0.05 & -0.1 & 0.05 \\ 0.1 & 0.2 & -0.3 \end{bmatrix}$ | | $\begin{bmatrix} -0.1 & 0.03 & 0.04 & 0.03 \\ 0.03 & -0.1 & 0.04 & 0.03 \\ 0.03 & 0.04 & -0.1 & 0.03 \\ 0.02 & 0.01 & 0.01 & -0.04 \end{bmatrix}$ |
| | $\begin{bmatrix} -0.1 & 0.1 \\ 0.8 & -0.8 \end{bmatrix}$ | | $\begin{bmatrix} -0.1 & 0.05 & 0.05 \\ 0.05 & -0.1 & 0.05 \\ 0.2 & 0.3 & -0.5 \end{bmatrix}$ | | $\begin{bmatrix} -0.1 & 0.03 & 0.04 & 0.03 \\ 0.03 & -0.1 & 0.04 & 0.03 \\ 0.03 & 0.04 & -0.1 & 0.03 \\ 0.02 & 0.02 & 0.01 & -0.05 \end{bmatrix}$ |
| | | | $\begin{bmatrix} -0.1 & 0.05 & 0.05 \\ 0.05 & -0.1 & 0.05 \\ 0.04 & 0.01 & -0.05 \end{bmatrix}$ | | $\begin{bmatrix} -0.1 & 0.03 & 0.04 & 0.03 \\ 0.03 & -0.1 & 0.04 & 0.03 \\ 0.03 & 0.04 & -0.1 & 0.03 \\ 0.06 & 0.07 & 0.07 & -0.2 \end{bmatrix}$ |
| | $\begin{bmatrix} -0.2 & 0.2 \\ 0.5 & -0.5 \end{bmatrix}$ | | $\begin{bmatrix} -0.3 & 0.1 & 0.2 \\ 0.1 & -0.3 & 0.2 \\ 0.05 & 0.05 & -0.1 \end{bmatrix}$ | | $\begin{bmatrix} -0.125 & 0.04 & 0.04 & 0.045 \\ 0.04 & -0.125 & 0.04 & 0.045 \\ 0.04 & 0.04 & -0.125 & 0.045 \\ 0.03 & 0.04 & 0.03 & -0.1 \end{bmatrix}$ |
| | | | $\begin{bmatrix} -0.3 & 0.1 & 0.2 \\ 0.1 & -0.3 & 0.2 \\ 0.1 & 0.2 & -0.3 \end{bmatrix}$ | | $\begin{bmatrix} -0.125 & 0.04 & 0.04 & 0.045 \\ 0.04 & -0.125 & 0.04 & 0.045 \\ 0.04 & 0.04 & -0.125 & 0.045 \\ 0.02 & 0.02 & 0.01 & -0.05 \end{bmatrix}$ |
| | $\begin{bmatrix} -0.2 & 0.2 \\ 0.8 & -0.8 \end{bmatrix}$ | | $\begin{bmatrix} -0.3 & 0.1 & 0.2 \\ 0.1 & -0.3 & 0.2 \\ 0.2 & 0.3 & -0.5 \end{bmatrix}$ | | $\begin{bmatrix} -0.125 & 0.04 & 0.04 & 0.045 \\ 0.04 & -0.125 & 0.04 & 0.045 \\ 0.04 & 0.04 & -0.125 & 0.045 \\ 0.02 & 0.01 & 0.01 & -0.04 \end{bmatrix}$ |
| | | | $\begin{bmatrix} -0.3 & 0.1 & 0.2 \\ 0.1 & -0.3 & 0.2 \\ 0.04 & 0.01 & -0.05 \end{bmatrix}$ | | $\begin{bmatrix} -0.125 & 0.04 & 0.04 & 0.045 \\ 0.04 & -0.125 & 0.04 & 0.045 \\ 0.04 & 0.04 & -0.125 & 0.045 \\ 0.06 & 0.07 & 0.07 & -0.2 \end{bmatrix}$ |
| | $\begin{bmatrix} -0.5 & 0.5 \\ 0.5 & -0.5 \end{bmatrix}$ | | $\begin{bmatrix} -0.5 & 0.2 & 0.3 \\ 0.2 & -0.5 & 0.3 \\ 0.05 & 0.05 & -0.1 \end{bmatrix}$ | | $\begin{bmatrix} -0.2 & 0.06 & 0.07 & 0.07 \\ 0.06 & -0.2 & 0.07 & 0.07 \\ 0.06 & 0.07 & -0.2 & 0.07 \\ 0.03 & 0.04 & 0.03 & -0.1 \end{bmatrix}$ |
| | | | $\begin{bmatrix} -0.5 & 0.2 & 0.3 \\ 0.2 & -0.5 & 0.3 \\ 0.1 & 0.2 & -0.3 \end{bmatrix}$ | | $\begin{bmatrix} -0.2 & 0.06 & 0.07 & 0.07 \\ 0.06 & -0.2 & 0.07 & 0.07 \\ 0.06 & 0.07 & -0.2 & 0.07 \\ 0.02 & 0.02 & 0.01 & -0.05 \end{bmatrix}$ |
| | $\begin{bmatrix} -0.5 & 0.5 \\ 0.8 & -0.8 \end{bmatrix}$ | | $\begin{bmatrix} -0.5 & 0.2 & 0.3 \\ 0.2 & -0.5 & 0.3 \\ 0.2 & 0.3 & -0.5 \end{bmatrix}$ | | $\begin{bmatrix} -0.2 & 0.06 & 0.07 & 0.07 \\ 0.06 & -0.2 & 0.07 & 0.07 \\ 0.06 & 0.07 & -0.2 & 0.07 \\ 0.02 & 0.01 & 0.01 & -0.04 \end{bmatrix}$ |

| | | |
|---|---|---|
| $\begin{bmatrix} -0.8 & 0.8 \\ 0.5 & -0.5 \end{bmatrix}$ | $\begin{bmatrix} -0.5 & 0.2 & 0.3 \\ 0.2 & -0.5 & 0.3 \\ 0.04 & 0.01 & -0.05 \end{bmatrix}$ | $\begin{bmatrix} -0.2 & 0.06 & 0.07 & 0.07 \\ 0.06 & -0.2 & 0.07 & 0.07 \\ 0.06 & 0.07 & -0.2 & 0.07 \\ 0.06 & 0.07 & 0.07 & -0.2 \end{bmatrix}$ |
| | $\begin{bmatrix} -0.05 & 0.04 & 0.01 \\ 0.04 & -0.05 & 0.01 \\ 0.05 & 0.05 & -0.1 \end{bmatrix}$ | $\begin{bmatrix} -0.5 & 0.2 & 0.2 & 0.1 \\ 0.2 & -0.5 & 0.2 & 0.1 \\ 0.2 & 0.2 & -0.5 & 0.1 \\ 0.03 & 0.04 & 0.03 & -0.1 \end{bmatrix}$ |
| | $\begin{bmatrix} -0.05 & 0.04 & 0.01 \\ 0.04 & -0.05 & 0.01 \\ 0.1 & 0.2 & -0.3 \end{bmatrix}$ | $\begin{bmatrix} -0.5 & 0.2 & 0.2 & 0.1 \\ 0.2 & -0.5 & 0.2 & 0.1 \\ 0.2 & 0.2 & -0.5 & 0.1 \\ 0.02 & 0.02 & 0.01 & -0.05 \end{bmatrix}$ |
| $\begin{bmatrix} -0.8 & 0.8 \\ 0.8 & -0.8 \end{bmatrix}$ | $\begin{bmatrix} -0.05 & 0.04 & 0.01 \\ 0.04 & -0.05 & 0.01 \\ 0.2 & 0.3 & -0.5 \end{bmatrix}$ | $\begin{bmatrix} -0.5 & 0.2 & 0.2 & 0.1 \\ 0.2 & -0.5 & 0.2 & 0.1 \\ 0.2 & 0.2 & -0.5 & 0.1 \\ 0.02 & 0.01 & 0.01 & -0.04 \end{bmatrix}$ |
| | $\begin{bmatrix} -0.05 & 0.04 & 0.01 \\ 0.04 & -0.05 & 0.01 \\ 0.04 & 0.01 & -0.05 \end{bmatrix}$ | $\begin{bmatrix} -0.5 & 0.2 & 0.2 & 0.1 \\ 0.2 & -0.5 & 0.2 & 0.1 \\ 0.2 & 0.2 & -0.5 & 0.1 \\ 0.06 & 0.07 & 0.07 & -0.2 \end{bmatrix}$ |

**Supporting Note 3. Generation of noise**

To evaluate the robustness of the denoising model under different noise characteristics, both Gaussian white noise and pink noise with varying intensities were added to the simulated signals. Gaussian white noise with a flat power spectral density (PSD) $S(f) = \sigma^2$, where $\sigma^2$ is the noise variance. For pink noise, the PSD follows $S(f) \propto 1/f$, where $f$ is the frequency. Gaussian white noise was used as a baseline noise model to represent random measurement noise commonly encountered in signal acquisition. In contrast, pink noise includes low-frequency components that are often present in experimental single-molecule measurements due to instrumental drift and environmental effects. By incorporating both noise types, the simulations cover a broader range of noise types, enabling a more comprehensive assessment of model performance under different conditions.

**Supporting Note 4. Optimization process**

To enhance the generalization capabilities of deep neural networks, we employed the *AdamW* optimizer during training with a weight decay of $1 \times 10^{-4}$. We also introduced the Cosine Annealing Scheduler to address the dual requirements of rapid convergence and refined search. The learning rate, as a hyperparameter, will be incorporated into subsequent search optimization. We introduced the cosine annealing scheduler to balance rapid convergence with finer search capabilities. The learning rate decays along a cosine curve based on epoch count, enabling the model to refine weights near local minima during late training phases while avoiding oscillations around optimal solutions. This learning rate will be incorporated as a hyperparameter into search optimization.

This study employed an automated Bayesian optimization strategy using the *Weights & Biases (W&B)* sweep framework. Our search objective is to maximize the *Score* value. This metric aims to find a balance point in model performance, maximizing state transition detection accuracy as indicated by a higher *F1*, while simultaneously minimizing signal reconstruction error as reflected by a lower *MSE*. Based on this objective function, the optimization algorithm performed 15 rounds of iterative searches within the predefined hyperparameter space, ultimately identifying the optimal parameter combination.

In each iteration, 5-fold cross-validation is performed on the current set of hyperparameters. This involves dividing the training data into five folds, training the model using four folds in each fold, calculating the score metric on the remaining fold, and finally aggregating the average Score across all folds as the score for that set of hyperparameters. The newly obtained hyperparameter combinations and their corresponding Score are fed back to the Bayesian optimization algorithm to update the model's understanding of the objective function.

The model was ultimately trained for 150 epochs with a batch size of 16. The initial learning rate is set to $7.61 \times 10^{-5}$ and is smoothly decayed to $1 \times 10^{-6}$ using cosine annealing strategy. The network is constructed based on 192 channels, employing channel scaling factors (1, 2, 4, 8) to build a 4-layer U-Net architecture, and incorporates a 4-head full-scale self-attention mechanism. In the final loss function, the amplitude weight $\lambda_{amp}$ = 14.53 and the edge weight $\lambda_{edge}$ = 8.95.

**Supporting Note 5. Design and implementation of low-pass filters**

We employed a 4-order low-pass filter to perform denoising on the signal. To more effectively compare the denoising performance of SSDM, we designed a parameter adaptation mechanism based on grid search to determine the optimal cutoff frequency, $f_s$. We have predefined certain key candidate cutoff frequencies based on the physical features of the signal (e.g., $f_s$ = [0.01, 0.08]). These values represent different levels of filtering intensity. Lower frequencies represent stronger smoothing effects, but may also smooth out signal details. Higher frequencies preserve more detail, but also retain more noise. Then the iterative phase begins, denoising with each candidate cutoff frequency in sequence. By evaluating the performance of each candidate cutoff frequency, we select the filter most suitable for the current signal. Regarding the evaluation, we employ the same method as SSDM. Using the same thresholding method to determine the transition points after denoising, we then calculate *F1*, *MSE*, and the *Score*. By comparing these results, we select the cutoff frequency with the highest *Score* as the cutoff frequency for the filter applied to the current signal.

**Supporting Note 6. Design and implementation of HMM**

We build an HMM model based on the BIC criterion to determine the number of signal states. However, since HMM is not a tool specifically designed for denoising, we adapt the HMM to better compare its denoising performance with that of SSDM. We define a candidate state set $S = \{2, 3, 4, 5, 6\}$, assuming the number of signal states belongs to the entire state set. The HMM model uses the number of states in each state to fit the current signal. During each fitting process, the HMM learns parameters corresponding to each state count, including the mean and state transition probabilities. We then utilize the BIC criterion to automatically select the optimal number of states. According to BIC = $-2\ln(L) + k\ln(n)$, where $L$ represents the maximum value of the likelihood function. The calculation of $\ln(L)$ employs the forward algorithm. This algorithm accumulates the sum of probabilities for all possible hidden state paths that generate the current observed data through dynamic programming. $k$ represents the total number of free parameters to be estimated in the model, and $n$ is the total number of data points observed. Ultimately, the model with the smallest BIC value will be selected as the optimal model. Then, the optimal model is used to predict the state for the entire signal segment. The model assigns the most probable hidden state to each time point and replaces the signal value at that point with the mean of the corresponding state, thereby obtaining the denoised signal. We employ the same evaluation method as SSDM. By identifying transition points in the signal using the thresholding method, we calculate *F1*, *MSE*, and the *Score*. This enables direct comparison of the denoising results with SSDM through the *Score* value.

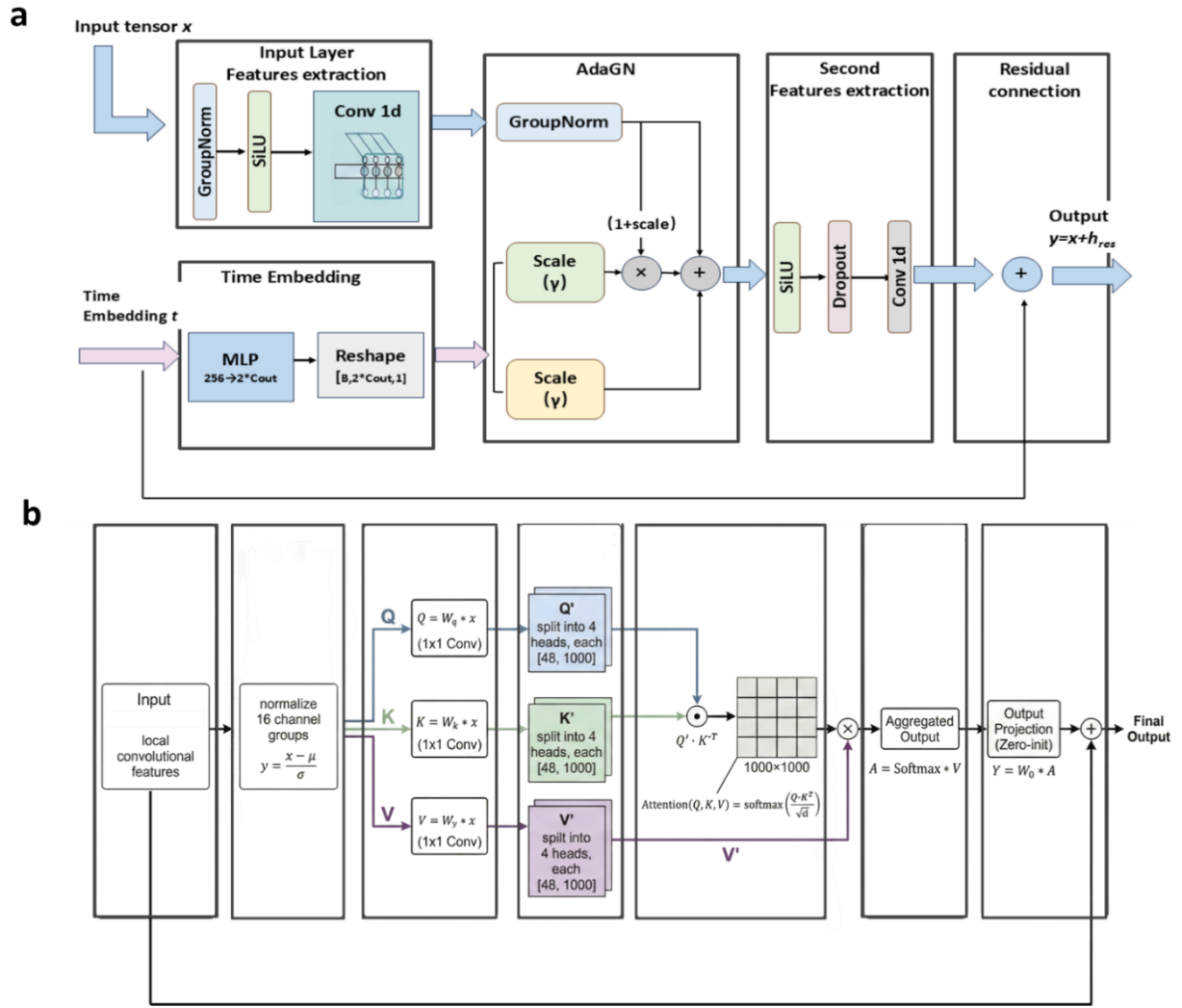

Figure S1. Architecture of the core components in the network. (a) Architecture of residual blocks in the network. (b) Architecture of attention blocks in the network.

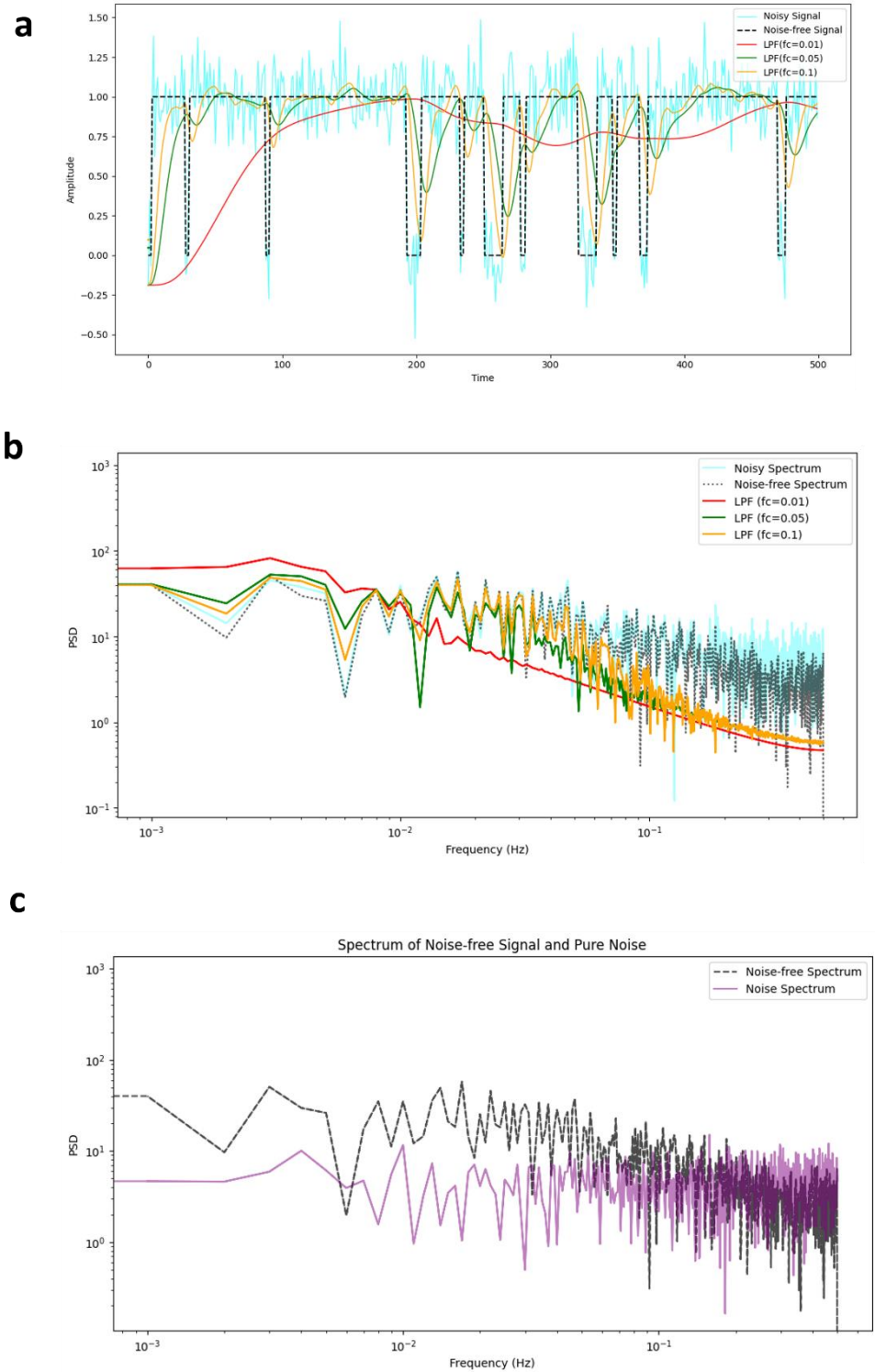

Figure S2. Comparison of processed signals by low-pass filters with different cutoff frequencies in time and frequency domains. (a) Time-domain plot of the noisy signal, the original noise-free signal, and the filtered outputs with cutoff frequencies $f_c$ = 0.01 Hz, 0.05 Hz, and 0.1 Hz. The sampling frequency of the signal is (normalized to) 1 Hz. (b) Spectra of the aforementioned signal traces. It can be clearly seen that the noise-free signal contains abundant high frequency components which is filtered out by low-pass filters. (c) Spectra of noise-free signal and pure noise. We can see that their spectra are highly overlapped, especially when $f$ > 0.1 Hz, which makes the directionless filtering impossible.

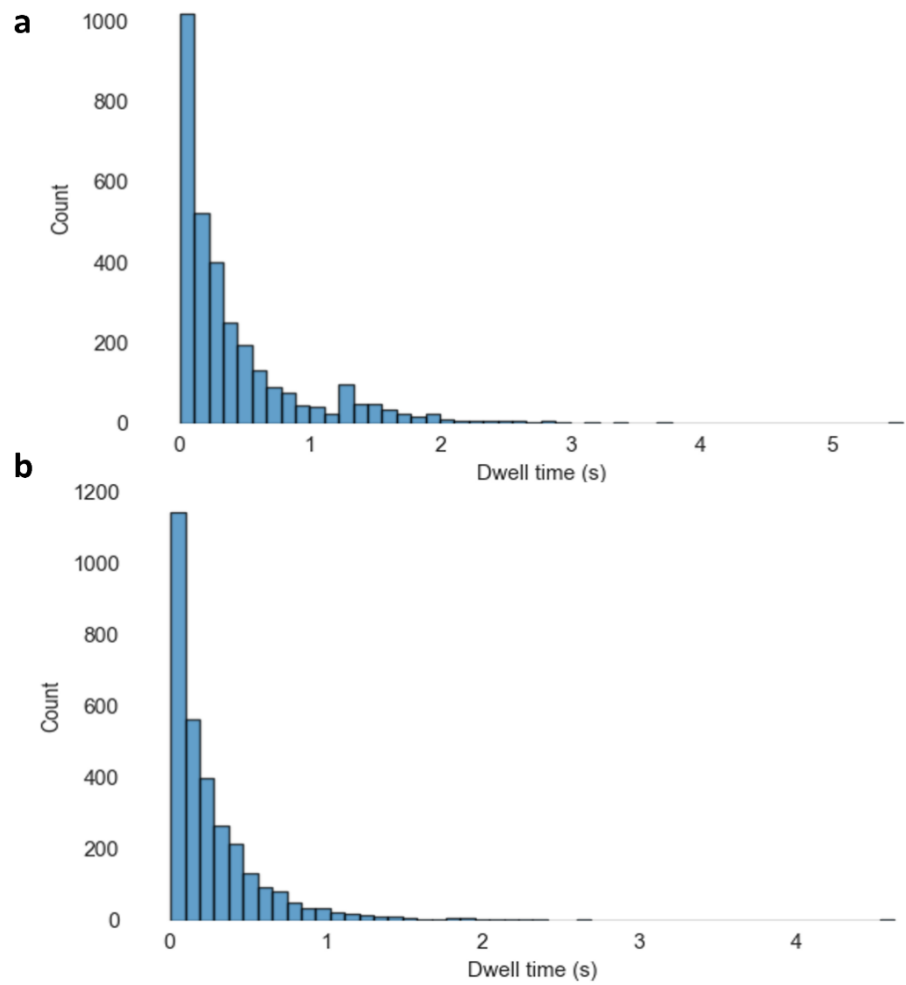

Figure S3. Histogram of dwell times of each state of the sm-FRET data. (a) Histogram of dwell times of state 1 in the sm-FRET signal. (b) Histogram of dwell times of state 2 in the sm-FRET signal.